\documentclass[journal]{IEEEtran}
\ifCLASSINFOpdf
\else
\fi

\usepackage{tikz}
\usepackage{amsmath}
\usepackage{amssymb}
\usepackage{graphics}
\usepackage{courier}
\usepackage{hyperref}
\usepackage{url}
\usepackage{algorithm}
\usepackage{color}
\usepackage[noend]{algpseudocode}

\usepackage{lineno}
\usepackage{ulem}

\makeatletter
\def\BState{\State\hskip-\ALG@thistlm}
\makeatother

\usepackage[justification=justified, font=footnotesize, labelfont=bf]{caption}

\begin{document}
%
\title{AdaNN: Adaptive Neural Network-based Equalizer via Online Semi-supervised Learning}
\normalem
\author{
	Qingyi~Zhou,~
	Fan~Zhang,~\IEEEmembership{Senior~Member,~IEEE,}
    and 		
	Chuanchuan~Yang,~\IEEEmembership{Senior~Member,~IEEE}%

\thanks{Manuscript received xx xx, xxxx; revised xx xx, xxxx \emph{(Corresponding author: Chuanchuan Yang)}.}  
\thanks{The authors are with the State Key Laboratory of Advanced Optical Communication Systems and Networks, Department of Electronics, Peking University, Beijing 100871, China (e-mail: zhouqingyi@pku.edu.cn, yangchuanchuan@pku.edu.cn, fzhang@pku.edu.cn). This work is funded by National Key R\&D Program of China under Grant 2018YFB1801702 and Joint Fund of the Ministry of Education under Grant 6141A02033347.}
}


%
%

\markboth{}
{Shell \MakeLowercase{\textit{et al.}}: Bare Demo of IEEEtran.cls for IEEE Journals}
%

\maketitle

\begin{abstract}
The demand for high speed data transmission has increased rapidly, leading to advanced optical communication techniques. 
In the past few years, multiple equalizers based on neural network (NN) have been proposed to recover signal from nonlinear distortions. 
However, previous experiments mainly focused on achieving low bit error rate (BER) on certain dataset with an offline-trained NN, neglecting the generalization ability of NN-based equalizer when the properties of optical link change. 
The development of efficient online training scheme is urgently needed. 
In this paper, we've proposed an adaptive online training scheme, which can fine-tune parameters of NN-based equalizer without the help of an online training sequence. 
By introducing data augmentation and virtual adversarial training, the convergence speed has been accelerated by 4.5 times, compared with decision-directed self-training. 
The proposed adaptive NN-based equalizer is called ``AdaNN''. 
Its BER has been evaluated under two scenarios: a 56 Gb/s PAM4-modulated VCSEL-MMF optical link (100-m), and a 32 Gbaud 16QAM-modulated Nyquist-WDM system (960-km SSMF). 
In our experiments, with the help of AdaNN, BER values can be quickly stabilized below 1e-3 after trained with $\text{10}^\text{5}$ unlabeled symbols. 
AdaNN shows great performance improvement compared with non-adaptive NN and conventional MLSE. 

\end{abstract}

\begin{IEEEkeywords}
Optical fiber communication, Adaptive nonlinear equalizer, Neural network, Semi-supervised learning.
\end{IEEEkeywords}

\IEEEpeerreviewmaketitle

\section{Introduction}
%
%
%
%


\IEEEPARstart{W}{ith} the continuous development of the Internet, higher bandwidth data transmission is required. Advanced modulation techniques together with novel algorithms have emerged to fulfill the requirements. Digital signal processing (DSP) is quite essential for improving the bit-error-rate (BER) performance and raising the optical link’s transmission rate. 

In order to achieve large transmission capacity in short-range optical interconnects, researchers have tried out a variety of conventional DSP techniques. With feed forward equalization (FFE), the data rate of non-return-to-zero (NRZ) has reached 71 Gb/s \cite{1}. Other conventional equalization techniques, such as decision feedback equalizer (DFE) and maximum likelihood sequence estimator (MLSE), have also been utilized \cite{2}-\cite{5}. By utilizing pre-emphasis, 94 Gb/s and 107 Gb/s PAM-4 transmission have been demonstrated by K. Szczerba et al. \cite{6} and J. Lavrencik et al. \cite{7} respectively. 

Researchers have also been trying to exploit the potential of DSP algorithms for long-reach optical communication systems. Volterra nonlinear equalizer (VNLE) has been utilized to mitigate nonlinear distortions \cite{8}. A few works have been done to lower the complexity of VNLE \cite{9, 10}. Other nonlinearity compensation techniques have also been investigated, such as digital back-propagation (DBP) \cite{11}, perturbation-based compensation \cite{12}, and nonlinear Kalman filter \cite{13}.

All the above-mentioned DSP algorithms are designed on rich expert knowledge, and some can be proved optimal for tractable mathematical models. However, many nonlinearities (modulation nonlinearity together with square law detection) that exist in practical systems can only be approximately captured and are difficult to compensate with conventional DSP techniques \cite{14}. In order to solve this problem, many DSP algorithms based on neural network have been proposed, including artificial neural network (ANN) based equalizer \cite{15, 16}, convolutional neural network (CNN) based equalizer \cite{17} and recurrent neural network (RNN) based equalizer \cite{18, 19}. Implemented in different optical communication systems, these NN-based equalizers have not only reached lower BER, but also shown excellent capability of mitigating nonlinearity. 

Although researchers report to have achieved lower BER using NN, there's one problem: it's difficult for NN to generalize over varied channel condition. In an actual communication system, the external environment and channel parameters may change, causing the distribution of received data to ``drift away''. For example, in data centers fibers are in motion due to rack vibration, which causes the channel properties to vary over time. An NN-based equalizer that performs well on training set/test set may suffer from severe performance degradation \cite{20}. On the other hand, it's too costly to train different NNs for different communication systems. Due to the lack of the ability to adjust parameters adaptively, existing NN-based equalizers cannot adapt to channel variations and thus, are not practical enough. 

Developing adaptive NN-based equalizer is therefore important. A new training scheme is expected, which does not rely on massive amount of collected labeled data. Some previous works on adaptive equalizers based on machine learning also require training sequence \cite{21, 22}. Unfortunately, similar parameter adjustment method cannot be used directly for NN-based equalizers. We've found that, when the short training sequence is provided to an NN-based equalizer, the equalizer still suffers from degraded BER performance. 
Researchers in the field of wireless communication are also exploring possible applications of deep learning techniques \cite{23}-\cite{27}. Most of these relevant works still rely on pilots when model parameters are changed adaptively \cite{23}. S. Schibisch et al. have used error-correcting codes (ECC) to construct labeled dataset for online training, but this causes overhead and relies on special protocol \cite{28}. In \cite{29}, the authors claimed that channel estimation based on semi-supervised learning is still an open subject.

In this paper, we propose an adaptive online training scheme, which can be used to fine-tune NN-based equalizer \textit{without} the help of training sequence. 
The proposed adaptive NN-based equalizer is called ``AdaNN''. 
The deployment of AdaNN include both offline training stage and online training stage. 
Although labeled training set is still required at offline stage, at online stage no labeled data needs to be provided. 
We collect recently received data using a sliding-window, then fine-tune parameters with the help of unlabeled data. 
Inspired by virtual adversarial training (VAT) which is a semi-supervised learning method, we propose a loss function named ``Aug-VAT'', which outperforms naive decision-directed self-training and leads to a 4.5 times speedup. 
AdaNN is evaluated under two scenarios: a 56 Gb/s PAM4-modulated short-distance (100-m) VCSEL-MMF optical interconnect system, and a 32 Gbaud 16QAM-modulated Nyquist-WDM system (960-km SSMF). 
Experimental results indicate that the BER performance of AdaNN is much better compared with non-adaptive NN and MLSE. 
Conclusions can be reached that without training sequence, it's possible to construct adaptive NN-based equalizer with acceptable computational cost, justifying the significance of our work.

The rest of this paper can be organized as follows. Section II provides a detailed introduction of our proposed online training scheme. In Section III, the computational complexity of proposed AdaNN is analyzed. In Section IV, the BER performance of AdaNN, non-adaptive NN, and MLSE are compared. Section V concludes the paper.

\section{AdaNN: Online Training based on Semi-supervised Learning}
\subsection{Nonlinear equalizer based on NN}
\begin{figure}[!t]
\centering
\includegraphics[width=0.48\textwidth]{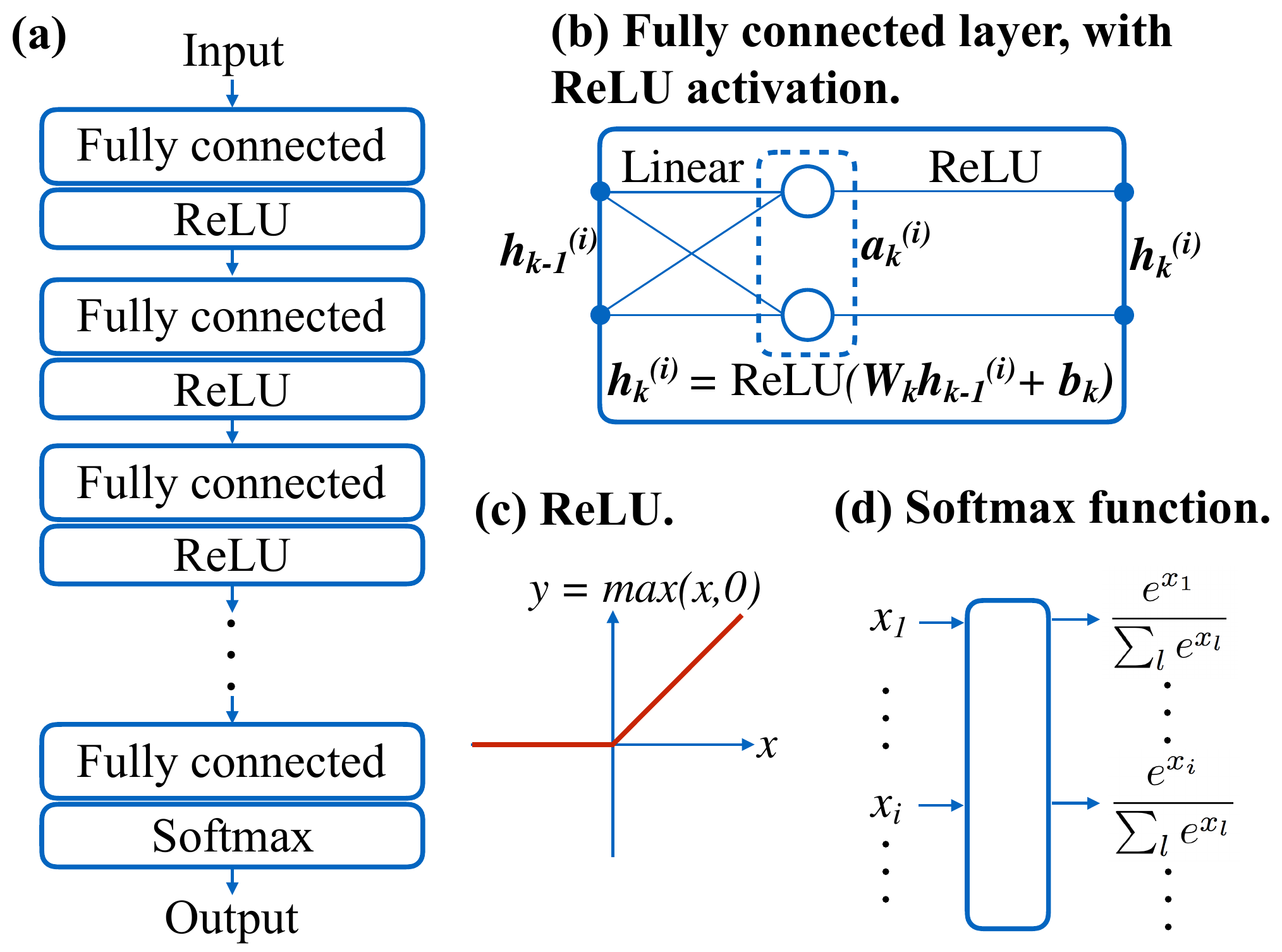}
\caption{(a) The structure of NN-based equalizer evaluated in this paper, including input layer, hidden layers and output layer. (b) Fully-connected layer, with ReLU activation. Note that only a few neurons are explicitly drawn. (c) The ReLU activation function. (d) The softmax function.}
\label{fig:structure}
\end{figure}
The NN we use contains an input layer, an output layer, and several hidden layers (each hidden layer contains $R$ neurons), as shown in Fig.~\ref{fig:structure}(a)(b). The total number of layers contained in this NN is denoted as $l_{\textrm{NN}}$. For the $i$-th symbol, the relationship between adjacent fully-connected layers (denoted as layer $k$ and $k-1$, where $k \in \{1, ..., l_{\textrm{NN}}-1 \}$) follows
\begin{equation}
\boldsymbol{a}^{(i)}_{k} = \boldsymbol{W}_{k}\boldsymbol{h}^{(i)}_{k-1} + \boldsymbol{b}_{l},
\label{eq:forward1}
\end{equation}
\begin{equation}
\boldsymbol{h}^{(i)}_{k} = \sigma(\boldsymbol{a}^{(i)}_{k}),
\label{eq:forward2}
\end{equation}
where $\boldsymbol{W}_{k}$ is $R \times R$ weight matrix, $\boldsymbol{b}_{k}$ is bias vector for layer $k$. Function $\sigma(\cdot)$ stands for activation function, with softmax chosen for the output layer and ReLU for hidden layers. Different activation functions are displayed in Fig.~\ref{fig:structure}(c)(d). 

\begin{figure*}[]
\centering
\includegraphics[width=0.70\textwidth]{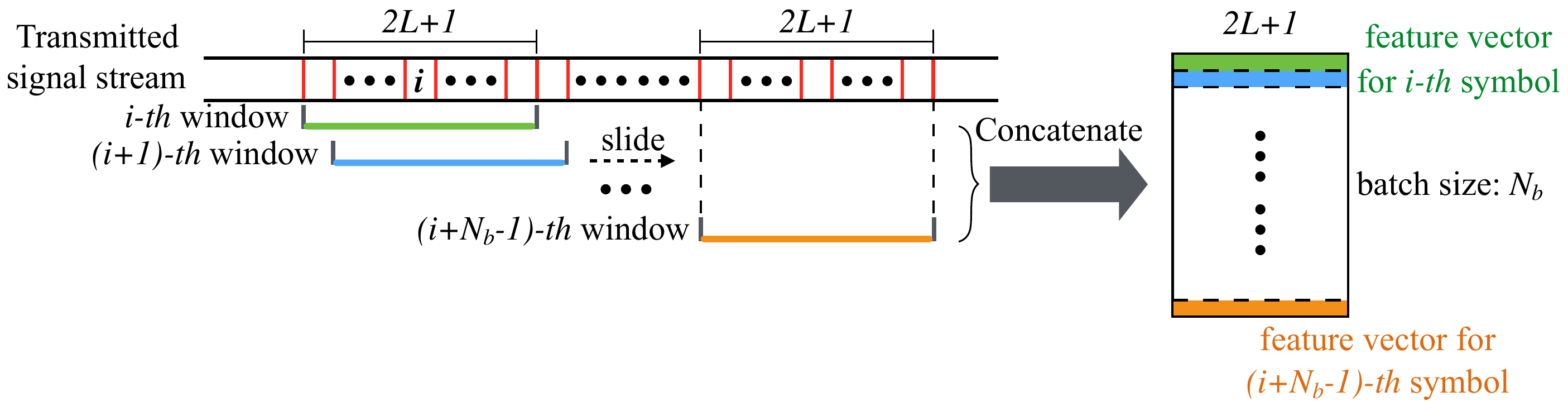}
\caption{A schematic illustration of the sliding window, which is used to collect batches. Parameters can be updated only if a batch (size $N_{b}$) has been collected.}
\label{fig:window}
\end{figure*}
\subsubsection{Offline Training Stage}
At the offline training stage, the loss function has the form of cross-entropy, which is widely used when dealing with multi-class classification \cite{30}. Denote the total number of symbols contained in the sequence as $N_{seq}$. The training process can be formulated as
\begin{equation}
\min_{\{ \boldsymbol{W}_{k}, \boldsymbol{b}_{k} \} } L_{loss} = \min_{ \{ \boldsymbol{W}_{k}, \boldsymbol{b}_{k} \}} (-\frac{1}{N_{seq}}\sum_{i=1}^{N_{seq}} \sum_{j=1}^{M} y_{j}^{(i)} \ln(o_{j}^{(i)})),
\label{eq:min}
\end{equation}
where $M$ means a symbol only belongs to one of $M$ classes. The loss function $L_{loss}$ measures the difference between predicted probability $\boldsymbol{o}^{(i)}$ and ground truth $\boldsymbol{y}^{(i)}$. The whole training dataset is divided into batches, each containing a small portion of all $N_{seq}$ training samples. The network parameters are updated iteratively using Stochastic Gradient Descent (SGD) optimizer with momentum, which is much faster compared with vanilla SGD \cite{30}. 

\subsubsection{Equalizing Process}
During equalization, we denote the received signal sequence after interpolation and zero-mean normalization as $\hat{\boldsymbol{r}} = [\hat{\boldsymbol{r}}_{1}, \hat{\boldsymbol{r}}_{2}, ..., \hat{\boldsymbol{r}}_{N_{seq}}]$, where vectors $\hat{\boldsymbol{r}}_{1}, ..., \hat{\boldsymbol{r}}_{N_{seq}}$ correspond to $N_{seq}$ received symbols (following chronological order). The feature vector $\boldsymbol{v}^{(i)}$ for the $i$-th symbol is constructed as
\begin{equation}
\boldsymbol{v}^{(i)} = [\hat{\boldsymbol{r}}_{i-L}, ..., \hat{\boldsymbol{r}}_{i}, ..., \hat{\boldsymbol{r}}_{i+L}].
\end{equation} 
We denote the interpolation multiple as $\Gamma$, thus the dimension of input feature vector $\boldsymbol{v}^{(i)}$ is $\Gamma (2L+1)$. The input-output relationship is given in Eq.~(\ref{eq:forward1})(\ref{eq:forward2}). 

\subsection{Proposed AdaNN online training scheme}
Suppose that we aim to classify symbols into $M$ classes correctly. For such a multi-class classification problem, data can be either ``labeled'' or ``unlabeled''. The term ``labeled'' means that for an input vector $\boldsymbol{x}^{(i)}$, the true label $\boldsymbol{y}^{(i)}$ (which is a one-hot vector) is provided. ``Unlabeled'' on the other hand, means that the exact classification result is not known. 

During online training stage, it is impossible to gather large amount of labeled data. It is possible that the transmitter provide short training sequences for channel estimation/parameter fine-tuning. Unfortunately, short training sequences are not enough for training NN. A possible solution is that, although the exact labels are not known, we can make use of the distribution of received signals to monitor the ``drift away'' process and use such information to fine-tune our equalizer. Here the concept of semi-supervised learning arises. Semi-supervised learning is a class of machine learning tasks that make use of unlabeled data for training (typically a small amount of labeled data with a large amount of unlabeled data). Unlabeled data helps us by providing information about the probability density distribution of input vectors \cite{31, 32}. 

Based on the idea of semi-supervised learning, we now explain the process of AdaNN. We focus on the online training stage in this part. First, during online stage a sliding window is utilized to collect data, which is illustrated in Fig.~\ref{fig:window}. The window, containing $2L+1$ symbols and denoted as bold line, slides on the received signal sequence. At each step $t$, $N_{b}$ input feature vectors $\boldsymbol{v}^{(1)}, ..., \boldsymbol{v}^{(N_{b})}$ are collected and serve as a batch. Gradient $\boldsymbol{g}_{t}$ is calculated based on the loss function, and parameters are updated. 

When no labels are provided, NN can work adaptively under decision-directed mode. However, if conventional cross-entropy loss function is used, the following problems occur: 

(1) When using vanilla cross-entropy loss function, the convergence speed is very slow. 

(2) In communication systems, signals are inevitably distorted by different levels of noise. Without data augmentation, the NN is not robust against noise. 

\begin{figure}[b]
\centering
\includegraphics[width=0.44\textwidth]{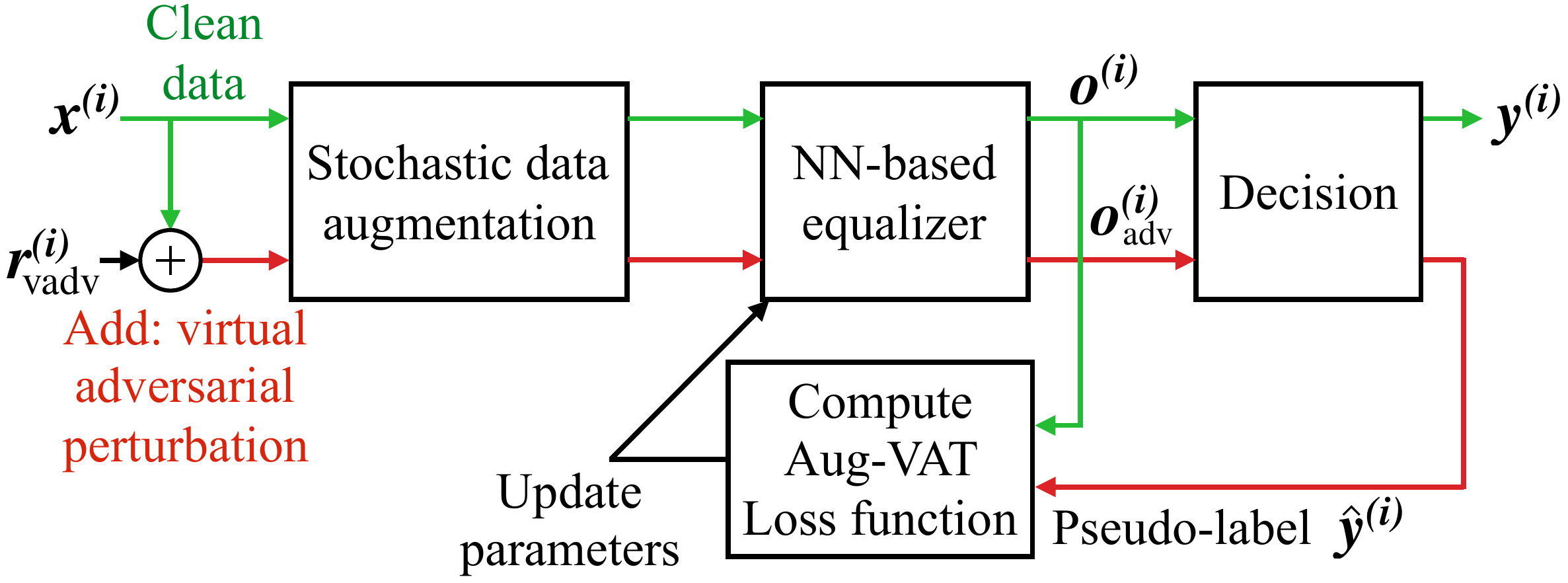}
\caption{The general structure of AdaNN, with Aug-VAT serves as loss function. AdaNN's parameters are updated with the help of pseudo-label $\hat{\boldsymbol{y}}^{(i)}$.}
\label{fig:pass}
\end{figure}

We've verified experimentally that both VAT and data augmentation can accelerate the training process greatly. Therefore, we propose a loss function named ``Augmented Virtual Adversarial Training'', or ``Aug-VAT'' for short. Aug-VAT combines $\Pi$-model \cite{33} and VAT \cite{34}, considering that the loss function should be consistent with the communication scenario. Fig.~\ref{fig:pass} shows the general structure of AdaNN with Aug-VAT. The detailed algorithm will be given as follows.

$\Pi$-model encourages consistent NN output between two realizations of one input vector, under two different data augmentation conditions. Denote $g_{\sigma}(\boldsymbol{v})$ as the input augmentation function. The augmentation is done by generating a random vector $\boldsymbol{\eta}$ and add it on $\boldsymbol{v}$:

\begin{equation}
g_{\sigma}(\boldsymbol{v}) = \boldsymbol{v} + \boldsymbol{\eta}, ~\boldsymbol{\eta} \sim \mathcal{N}(\boldsymbol{0}, \sigma^{2}\boldsymbol{I}_{\Gamma (2L+1)}).
\label{eq:g_sigma}
\end{equation}

When using Aug-VAT as loss function in AdaNN, every single input feature vector $\boldsymbol{v}$ should first be replaced using $g_{\sigma}(\boldsymbol{v})$, then serve as the input feature vector in VAT. VAT is closely related to adversarial training \cite{35}. The adversarial perturbation for the $i$-th input vector can be defined as
\begin{equation}
\boldsymbol{r}^{(i)}_{\textrm{adv}} = \arg \max_{\boldsymbol{r};\| \boldsymbol{r} \| \leqslant \epsilon} \{ -\sum_{j=1}^{M} y_{j}^{(i)} \ln[ (\textrm{NN}_{\boldsymbol{\theta}}(g_{\sigma}(\boldsymbol{v}^{(i)} + \boldsymbol{r})))_{j}]\}.
\end{equation}
This equation implies that by adding a small perturbation $\boldsymbol{r}^{(i)}_{\textrm{adv}}$ (satisfying $\| \boldsymbol{r}^{(i)}_{\textrm{adv}} \| \leqslant \epsilon$) on $\boldsymbol{v}^{(i)}$, the loss function calculated using the perturbed input tend to increase. ``Adversarial training'' means that during training the loss function is always calculated based on the perturbed input vectors rather than the clean ones, so that NN's robustness can be improved. When full label information $\boldsymbol{y}^{(i)}$ is not available, $\boldsymbol{r}_{\textrm{adv}}$ can only be approximated by computing $\boldsymbol{r}_{\textrm{vadv}}$, which is derived efficiently using one-time power iteration method (see Algorithm \ref{alg:radv}).

The complete form of Aug-VAT loss function for a single batch can be formulated as
\begin{equation}
\begin{split}
L_{loss} & = -\frac{1}{N_{b}}\sum_{i=1}^{N_{b}} \sum_{j=1}^{M} \hat{y}_{j}^{(i)} \ln[( \textrm{NN}_{\boldsymbol{\theta}}( g_{\sigma}(\boldsymbol{v}^{(i)}) )  )_{j}] \\
& = -\frac{1}{N_{b}}\sum_{i=1}^{N_{b}} \hat{y}_{l_{\textrm{adv}}}^{(i)} \ln[( \textrm{NN}_{\boldsymbol{\theta}}( g_{\sigma}(\boldsymbol{v}^{(i)}) )  )_{l_{\textrm{adv}}}],
\end{split}
\end{equation}
where index $l_{\textrm{adv}}$ means that after adding adversarial perturbation and noise, the $i$-th symbol is classified into class $l_{\textrm{adv}}$:
\begin{equation}
l_{\textrm{adv}} = \arg \max_{k} [ \textrm{NN}_{\boldsymbol{\theta}} (g_{\sigma} (\boldsymbol{v}^{(i)} + \boldsymbol{r}^{(i)}_{\textrm{vadv}}) ) ]_{k}.
\end{equation}

\begin{algorithm}
\caption{Virtual adversarial perturbation}
\label{alg:radv}
\begin{algorithmic}[1]
\Require $\boldsymbol{v}^{(i)}$ = $i$-th input feature vector ($i \in \mathbb{Z}^{+}$)
\Require $\boldsymbol{\theta}$ = parameters of the offline-trained network
\Require $\xi$ = step size of gradient estimation (default: $0.1$)
\Require $\epsilon$ = length of adversarial perturbation

\State Generate random vector $\boldsymbol{d}^{(i)}$ with Gaussian distribution.
\State $\boldsymbol{r} \gets \xi \boldsymbol{d}^{(i)}$
\State $\boldsymbol{o}^{(i)} \gets \textrm{NN}_{\boldsymbol{\theta}}(g_{\sigma}(\boldsymbol{v}^{(i)}))$
\State $\boldsymbol{o'}^{(i)} \gets \textrm{NN}_{\boldsymbol{\theta}}(g_{\sigma}(\boldsymbol{v}^{(i)} + \boldsymbol{r}))$
\State Compute $L_{loss} = D_{KL}(\boldsymbol{o}^{(i)} \| \boldsymbol{o'}^{(i)}) = \sum_{j=1}^{M}o^{(i)}_{j} \ln (\frac{o^{(i)}_{j}}{{o'}^{(i)}_{j}})$ 
$\triangleright$ Difference between two output vectors can be quantified using Kullback-Leibler (KL) divergence
\State $\boldsymbol{g}^{(i)} \gets \nabla_{\boldsymbol{r}} L_{loss}$
\State $\boldsymbol{r}^{(i)}_{\textrm{vadv}} \gets \boldsymbol{g}^{(i)}/ \| \boldsymbol{g}^{(i)}\|$
\State $\boldsymbol{r}^{(i)}_{\textrm{vadv}} \gets \epsilon \cdot \boldsymbol{r}^{(i)}_{\textrm{vadv}}$

\end{algorithmic}
\end{algorithm}

\begin{figure*}[]
\centering
\includegraphics[width=0.9\textwidth]{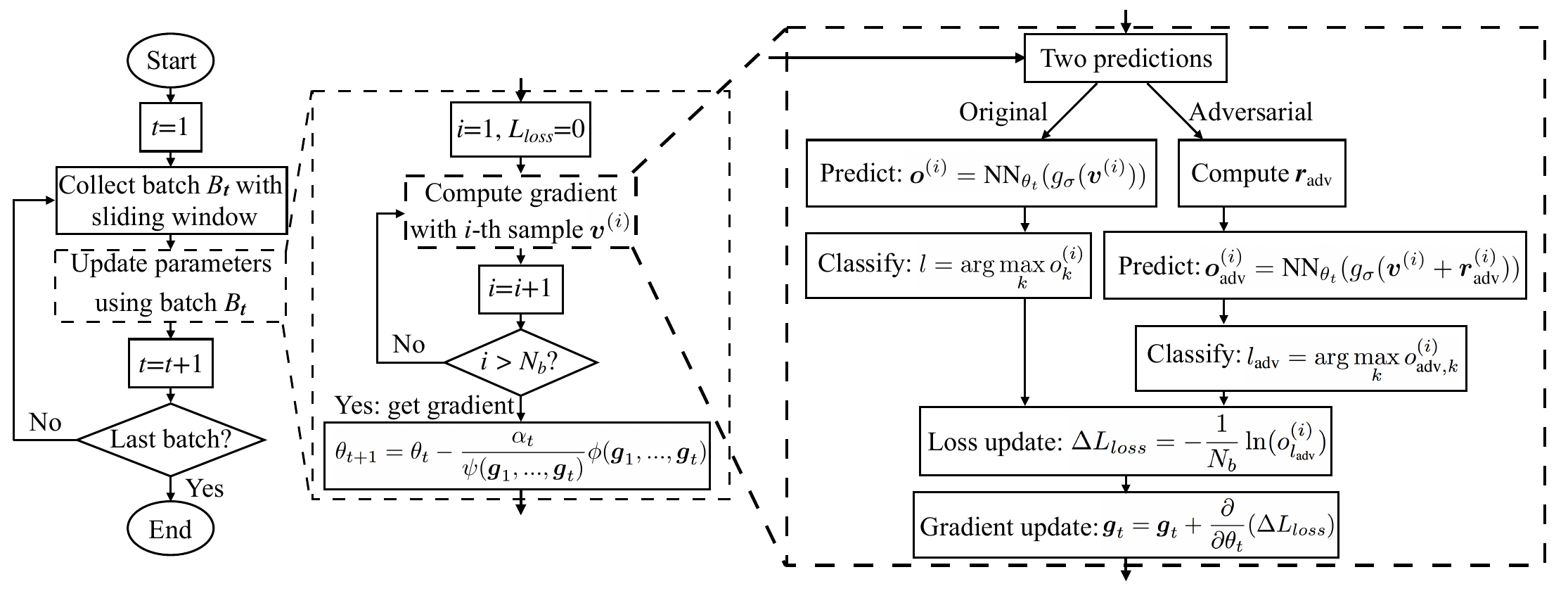}
\caption{The complete flow chart of the AdaNN process. The box with thinner dashed lines represents the processing of a single batch. The box with thicker dashed lines, on the other hand, represents the processing of the $i$-th sample in the batch.}
\label{fig:process}
\end{figure*}

The final pseudocode of our proposed AdaNN online training scheme (with Aug-VAT as loss function) is given in Algorithm \ref{alg:AdaNN}. The flow chart of AdaNN is displayed in Fig.~\ref{fig:process}. At step $t$, the gradient $\boldsymbol{g}_{t}$ is accumulated before all the data in batch $B_{t}$ have been utilized. After that, parameters $\boldsymbol{\theta}_{t}$ should be updated using gradient-based optimizer. Here, all gradient-based optimization algorithms can be written in the following general form \cite{36}:
\begin{equation}
\boldsymbol{\theta}_{t+1} = \boldsymbol{\theta}_{t} - \frac{\alpha_{t}}{\psi(\boldsymbol{g}_{1}, ..., \boldsymbol{g}_{t})}\phi(\boldsymbol{g}_{1},..., \boldsymbol{g}_{t}),
\label{eq:general_optimizer}
\end{equation}
where $\boldsymbol{g}_{t}$ represents the gradient obtained at the $t$-th time step, $\alpha_{t}/\psi(\boldsymbol{g}_{1}, ..., \boldsymbol{g}_{t})$ denotes the adaptive learning rate, and $\phi(\boldsymbol{g}_{1},..., \boldsymbol{g}_{t})$ is the gradient estimation. Several influential optimizers include: SGD \cite{37}, Momentum SGD \cite{38}, Nesterov Momentum \cite{39}, AdaGrad \cite{40}, RMSprop \cite{30}, and Adam \cite{41}. Choosing the right optimizer has great impact on AdaNN's performance. Experimental results show that Adam performs the best for our task.
\begin{algorithm}
\caption{AdaNN}
\label{alg:AdaNN}
\begin{algorithmic}[1]
\Require $\boldsymbol{v}^{(i)}$ = $i$-th input feature vector ($i \in \mathbb{Z}^{+}$)
\Require $\boldsymbol{\theta}_{1}$ = parameters of the offline-trained network
\Require $g_{\sigma}(\cdot)$ = add Gaussian noise with deviation $\sigma$

\State \textbf{for} ${t = 1,2,3...}$ \textbf{do}
	\State \quad \textbf{for} $i$ in [1, $N_{b}$] \textbf{do}
		\State \quad \quad $\boldsymbol{o}^{(i)} \gets \textrm{NN}_{\boldsymbol{\theta}_{t}}(g_{\sigma}(\boldsymbol{v}^{(i)}))$
		\State \quad \quad $l \gets \arg \max_{k} o^{(i)}_{k}$ \quad $\triangleright$ predicted label
		\State \quad \quad compute $\boldsymbol{r}^{(i)}_{\textrm{vadv}}$ using Algorithm \ref{alg:radv}
		\State \quad \quad $\boldsymbol{o}_{\textrm{adv}}^{(i)} \gets \textrm{NN}_{\boldsymbol{\theta}_{t}}(g_{\sigma}(\boldsymbol{v}^{(i)} + \boldsymbol{r}^{(i)}_{\textrm{vadv}}))$
		\State \quad \quad $l_{\textrm{adv}} \gets \arg \max_{k} o^{(i)}_{\textrm{adv}, k}$ \quad $\triangleright$ perturbed label
		\State \quad \quad $\hat{y}_{l_{\textrm{adv}}}^{(i)} = 1$ \quad $\triangleright$ one-hot perturbed label vector
		\State \quad \quad $L_{loss} \gets L_{loss} - \frac{1}{N_{b}}\sum_{j=1}^{M} \hat{y}_{j}^{(i)} \ln(o_{j}^{(i)})$
	
	\State \quad \textbf{end for}
	
	\State \quad $\boldsymbol{g}_{t} \gets \frac{\partial L_{loss}}{\partial \boldsymbol{\theta}_{t}}$ \quad $\triangleright$ compute gradient on batch
	\State \quad update $\boldsymbol{\theta}_{t}$ using gradient-based optimizer (e.g., Adam)
	
\State \textbf{end for}

\end{algorithmic}
\end{algorithm}

\subsection{Other choices for loss function}
When labels are not provided,  $y_{j}^{(i)}$ in Eq.~(\ref{eq:min}) should be replaced with pseudo-label $\hat{y}_{j}^{(i)}$. The loss function still has the cross-entropy form
\begin{equation}
L_{loss} = -\frac{1}{N_{b}}\sum_{i=1}^{N_{b}} \sum_{j=1}^{M} \hat{y}_{j}^{(i)} \ln(o_{j}^{(i)}).
\end{equation}
Pseudo-label $\hat{y}_{j}^{(i)}$ can be obtained in different ways, corresponding to different loss functions. In the last subsection, we've proposed Aug-VAT as loss function, which combines $\Pi$-model with VAT. Besides, vanilla self-training, $\Pi$-model and VAT can also be used alone as loss function. 

(1) Self-training: For the $i$-th input feature vector $\boldsymbol{v}^{(i)}$, the output probability vector $\boldsymbol{o}^{(i)}=\textrm{NN}_{\boldsymbol{\theta}}(\boldsymbol{v}^{(i)})$. The pseudo-label $\hat{\boldsymbol{y}}^{(i)}$ can be derived by
\begin{equation}
\hat{y}_{j}^{(i)}=
\left\{\begin{matrix}
1,~\textrm{if}~j=\arg \max_{k}(o_{k}^{(i)}), \\ 
0,~\textrm{otherwise.} 
\end{matrix}\right.
\label{eq:one-hot}
\end{equation}
Self-training is similar to decision-directed mode of conventional adaptive equalizers, and thus serves as a baseline. 
\begin{figure*}[t]
\centering
\includegraphics[width=0.65\textwidth]{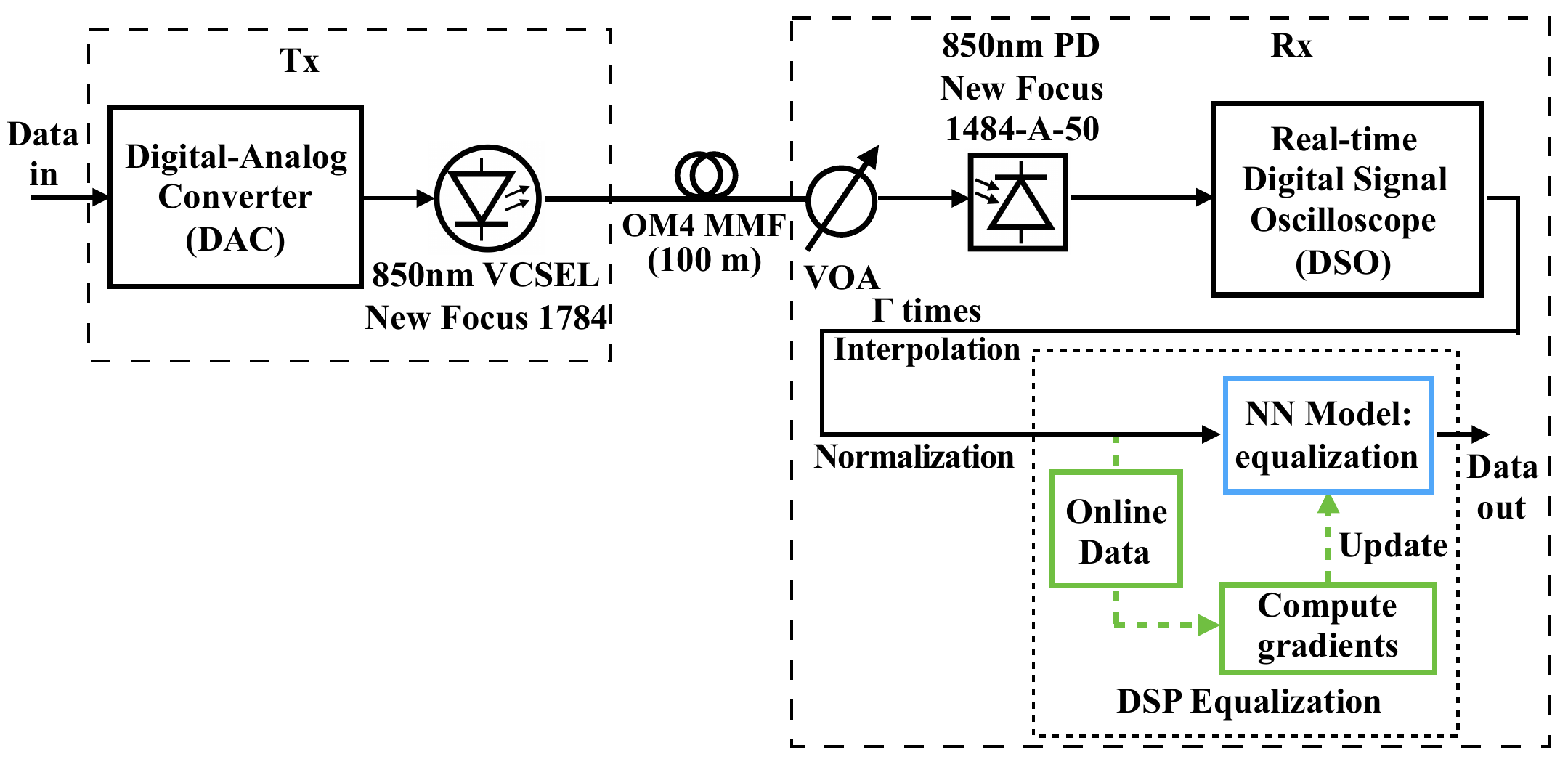}
\caption{Experimental configuration of the 56 Gb/s PAM-4 signal transmission system utilizing 850-nm VCSEL and OM4 MMF.}
\label{fig:system}
\end{figure*}

(2) $\Pi$-model only: The main difference between $\Pi$-model and self-training lies in data augmentation. The output probability vector $\boldsymbol{o}^{(i)}=\textrm{NN}_{\boldsymbol{\theta}}(g_{\sigma}(\boldsymbol{v}^{(i)}))$, where $g_{\sigma}(\cdot)$ follows Eq.~(\ref{eq:g_sigma}). The derivation of $\hat{\boldsymbol{y}}^{(i)}$ is the same as Eq.~(\ref{eq:one-hot}). 

(3) Virtual adversarial training only: When using vanilla VAT, the output probability vector $\boldsymbol{o}_{\textrm{adv}}^{(i)}=\textrm{NN}_{\boldsymbol{\theta}}(\boldsymbol{v}^{(i)} + \boldsymbol{r}_{\textrm{vadv}}^{(i)})$, where $\boldsymbol{r}_{\textrm{vadv}}^{(i)}$ is the virtual adversarial perturbation vector calculated from Algorithm \ref{alg:radv}. The derivation of $\hat{\boldsymbol{y}}^{(i)}$ is the same as Eq.~(\ref{eq:one-hot}). 

All these loss functions are compared in Section IV. There are several other loss functions we haven't covered. ``Temporal ensemble'' \cite{33} requires re-evaluation of all training samples each time the NN parameters are updated, which is too costly. ``Mean teacher'' \cite{42} constructs an ensemble using current model and several past models during training. Our experiments show that ``Mean teacher'' has no difference compared with self-training. We also know that many semi-supervised learning algorithms are based on ``low dimension manifold assumption'', which assumes that data lie on a manifold of much lower dimension compared with input space. Relevant algorithms include low dimension manifold model (LDMM) and curvature regularization (CURE) \cite{43, 44}. However the estimation of local dimension/curvature requires access to all data points in a small area, which cannot be guaranteed. 

\section{Computational Complexity}
In this section, we focus on analyzing the computational cost of AdaNN. Fully-connected NNs mainly involve two types of computations: multiplications and activation functions. Note that here rather than $\tanh(\cdot)$, ReLU activation function is used. Therefore when analyzing complexity, activation doesn't need to be considered. For a non-adaptive deep neural network, the calculation of output probability vector $\boldsymbol{o}$ is called ``forward propagation'', which follows Eq.~(\ref{eq:forward1})(\ref{eq:forward2}). When equalizing a single symbol, each layer can be viewed as a vector. The number of neurons contained in all $l_{\textrm{NN}}$ layers are: $\Gamma (2L+1)$, $R$, $R$, ..., $R$, and $M$. The number of floating-point multiplications $k_{\textrm{NN}}$ can be calculated as
\begin{equation}
k_{\textrm{NN}} = \Gamma(2L+1)\cdot R + (l_{\textrm{NN}}-3)\cdot R^{2} + R\cdot M.
\end{equation}

As for AdaNN, all the parameters need to be adjusted online. According to Appendix. A, for a single back-propagation, the number of required floating-point multiplications $k_{\textrm{back}}$ can be calculated as
\begin{equation}
k_{\textrm{back}} = 2k_{\textrm{NN}} + (l_{\textrm{NN}}-2)\cdot R + M \approx 2k_{\textrm{NN}}.
\end{equation}
The computational cost of back-propagation is slightly larger than two times the cost of forward propagation. 

When $\Pi$-model serves as loss function, two forward propagations and one back-propagation are needed in a single iteration. Thus the computational cost of AdaNN ($\Pi$-model as loss function) should be approximately 4 times the cost of non-adaptive NN. When Aug-VAT serves as loss function, two forward propagations, one back-propagation and computing $\boldsymbol{r}_\textrm{vadv}$ (mainly includes one forward propagation and one back-propagation) are needed in a single iteration. In total, the computational cost of AdaNN (Aug-VAT as loss function) should be approximately 7 times the cost of non-adaptive NN. As a contrast, the computational cost of AdaNN (self-training as loss function) is 3 times the cost of non-adaptive NN. 

\section{Experimental Results}
In order to justify AdaNN's wide applicability, we've conducted experiments under two scenarios: a 56 Gb/s PAM4-modulated VCSEL-MMF optical link (100-m), and a 32 Gbaud 16QAM-modulated Nyquist-WDM system (960-km SSMF). The results are analyzed in this section. 

\subsection{Different Loss Functions}
We first present the adaptive training results for all the $4$ different loss functions. In this subsection experiments are conducted with a 56 Gb/s PAM4-modulated VCSEL optical link, which is depicted in Fig.~\ref{fig:system}. The system mainly consists of a directly modulated 850-nm VCSEL, 100-m OM4 MMF, and a photodiode (PD). The received signal is sampled using a high-speed real-time digital signal oscilloscope (DSO). The 850-nm VCSEL is New Focus$^{\circledR}$ 1784, while PD is New Focus$^{\circledR}$ 1484-A-50. The OM4 MMF is chosen as YOFC$^{\circledR}$ MaxBand$^{\circledR}$ OM4 multimode fiber. The DSO is Agilent DSAX96204Q, with sampling rate of $160$ GSa/s. We first 4x resample the received signal as stated in \cite{45} ($\Gamma=4$ for all experiments). The signal is then normalized, and input feature vectors are constructed. We've generated two sets of PAM-4 symbols with Bit-pattern Generator (BPG) of SHF 12104A (56 Gb/s). Following \cite{46}, we did not use PRBS pattern. Instead, a binary sequence is first generated by applying $\textrm{sign}(\cdot)$ function to an Gaussian noise sequence generated in MATLAB, then converted into two PAM-4 sequences. Each of the two datasets (denoted as set1 and set2) contains $2^{20}$ PAM-4 symbols. Set2 was collected $56$ hours after we collected set1. For both set1 and set2, the receive optical power (ROP) is $-2.7$ dBm. Between these two experiments, we rebuilt the experimental system and adjusted the position of optical fiber, in order to simulate a realistic scenario where fiber properties change slightly. Our NN-based equalizer with $4$ hidden layers ($l_{\textrm{NN}}=6$, $R=10$) is first trained offline using $25\%$ data in set1. The tap number of NN is first optimized by testing $L \in \{ 1, 3, 5, 7, 9, 11\}$. Then tap number is fixed as $L=5$ since it achieves satisfying BER performance. The batch size is fixed as $N_{b} = 16384$ for offline training. Momentum SGD is used, with initial learning rate $\alpha=0.004$ and moving average decay $\beta=0.9$. The model is trained for $200$ epochs (An epoch represents a single pass through the entire training set, meaning that all feature vectors in the training set have been used for exactly one time). This ensures good convergence and a BER lower than $10^{-3}$.

During online stage, a sliding window is utilized to collect data, as Fig.~\ref{fig:window} shows. Set1 and set2 are concatenated and then processed sequentially. The batch size for online training is $N_b = 8192$. Adam optimizer is used, with initial learning rate $\alpha = 0.01$, $\beta_{1} = 0.9$, and $\beta_{2} = 0.999$. When processing set1 and set2 sequentially, the BER for set1 will remain relatively low, while for set2 the BER will increase abruptly. By utilizing an online training scheme, hopefully the BER will then decrease to a low level. BER curves obtained are smoothed by averaging the BER values of neighboring $8$ batches. We mainly focus on two quantities: 

(1) Convergence time, defined as the number of batches it takes before AdaNN satisfies two conditions: reaching a BER lower than $10^{-3}$ on recent $8$ batches, and reaching an overall BER lower than $10^{-3}$ on set2. 

(2) Final BER, defined as AdaNN's BER on set2 at the end of online training stage. 

The convergence time as well as final BER are summarized in Table.~\ref{tab:BER_loss}. The numbers in bold represents the best performance among one class of training method (only hyper-parameters are changed). From Table.~\ref{tab:BER_loss} we can tell that, while self-training suffers from slow convergence, AdaNN can be $4.5$ times faster ($72 \div 16 = 4.5$), which indicates AdaNN's effective usage of unlabeled data. The final BER values on set2 are also displayed in order to show AdaNN's good generalization ability. In the following experiments, AdaNN's loss function is fixed as Aug-VAT ($\sigma=0.15$, $\epsilon = 0.3$). We've also calculated the change of weight matrices before and after the online training stage. The results are given in Appendix. B.

\begin{table}[!h]
	\caption{Convergence time and final BER performance (set2) of AdaNN, with different loss functions, including self-training, $\Pi$-model, VAT, and Aug-VAT. The $95\%$ confidence interval of BER estimations are also provided.}
	\label{tab:BER_loss}
	
	\centering
	\begin{tabular}{cccc}
	    \hline
		\hline
		Loss & Converge & Final BER & $95\%$ confidence \\
		function & time (batch) & ($\times10^{-4}$) & interval ($\times10^{-4}$) \\
		\hline
		\hline
		Self-training & $72$ & $8.39$ & $[6.32, 10.84]$ \\
		\hline
		$\Pi$-model ($\sigma$=0.1) & $24$ & $4.27$ & $[2.84, 6.08]$ \\
		$\Pi$-model ($\sigma$=0.2) & $\boldsymbol{16}$ & $2.75$ & $[1.63, 4.25]$ \\
		$\Pi$-model ($\sigma$=0.3) & $\boldsymbol{16}$ & $2.44$ & $[1.67, 3.45]$ \\
		$\Pi$-model ($\sigma$=0.4) & $32$ & $3.43$ & $[2.16, 5.08]$ \\
		\hline
		VAT ($\epsilon$=0.1) & $40$ & $5.34$ & $[3.72, 7.34]$ \\
		VAT ($\epsilon$=0.2) & $24$ & $2.98$ & $[1.80, 4.53]$ \\
		VAT ($\epsilon$=0.3) & $\boldsymbol{16}$ & $2.44$ & $[1.67, 3.45]$ \\
		VAT ($\epsilon$=0.4) & $\boldsymbol{16}$ & $2.52$ & $[1.45, 3.96]$ \\
		VAT ($\epsilon$=0.5) & $24$ & $2.82$ & $[1.69, 4.34]$ \\
		\hline
		Aug-VAT ($\sigma$=0.10,$\epsilon$=0.3) & $\boldsymbol{16}$ & $2.59$ & $[1.51, 4.06]$ \\
		Aug-VAT ($\sigma$=0.15,$\epsilon$=0.3) & $\boldsymbol{16}$ & $2.29$ & $[1.28, 3.68]$ \\
		Aug-VAT ($\sigma$=0.20,$\epsilon$=0.3) & $\boldsymbol{16}$ & $2.75$ & $[1.63, 4.25]$ \\
		Aug-VAT ($\sigma$=0.30,$\epsilon$=0.3) & $\boldsymbol{16}$ & $3.36$ & $[2.10, 4.99]$ \\
		\hline
		\hline
		
	\end{tabular}
\end{table}

\subsection{100-m VCSEL-MMF link}
In this subsection, AdaNN is evaluated in the 100-m VCSEL-MMF optical link described above. During online stage, set1 and set2, each containing $128$ batches (batch size $N_{b}=8192$) are concatenated and processed sequentially. The BER curve of AdaNN is compared with multiple equalizers, including non-adaptive NN, NN with training sequence, and MLSE. 
\subsubsection{Compare: Non-Adaptive NN}
\begin{figure}[!b]
\centering
\includegraphics[width=0.43\textwidth]{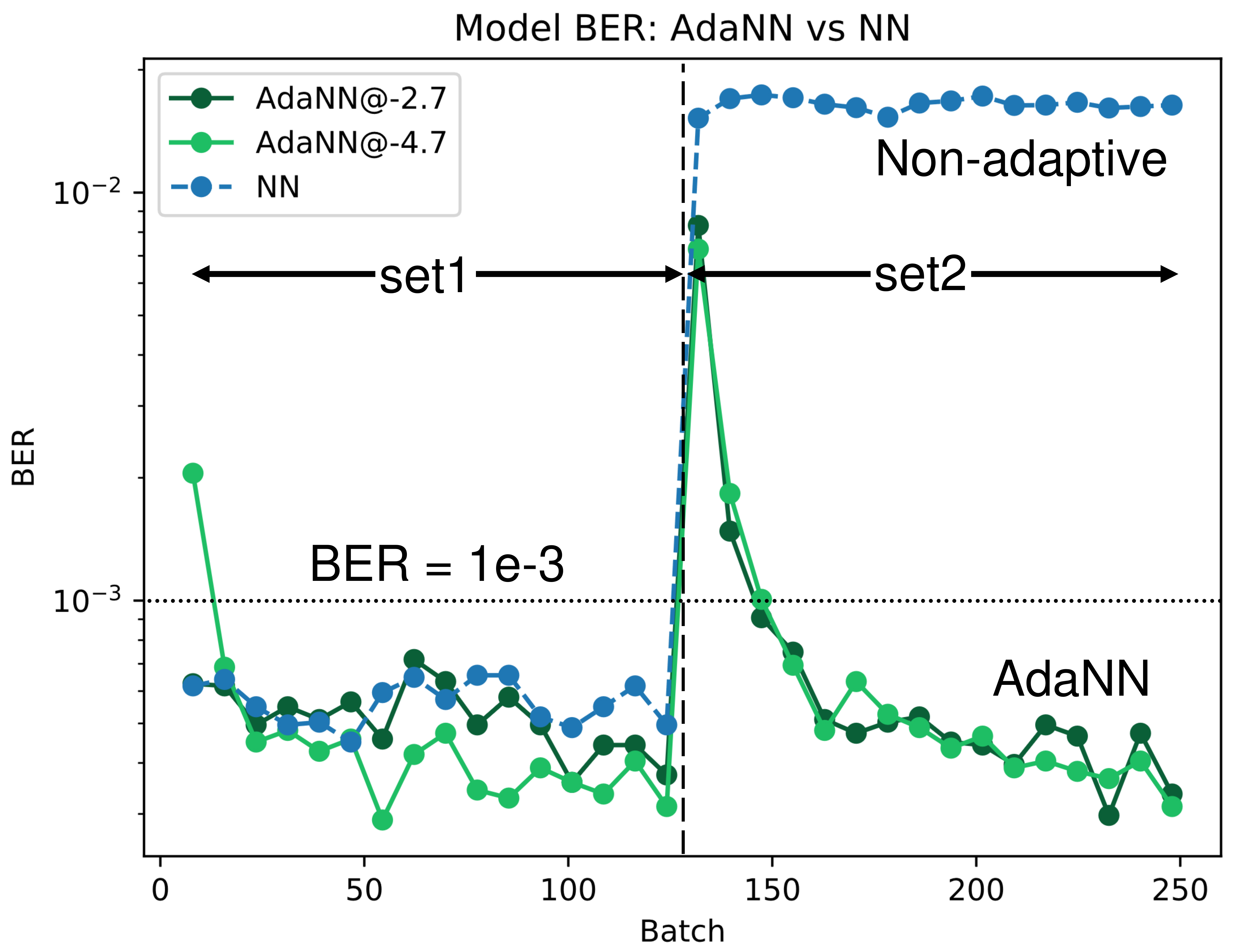}
\caption{The BER performance of a normal NN-based equalizer, compared with AdaNN. Note that performance of AdaNN trained with a different dataset (received optical power is $-4.7$ dBm) is also plotted, in order to show AdaNN's robustness.}
\label{fig:AdaNN_vs_NN}
\end{figure}

The BER performance of AdaNN as well as a non-adaptive NN is displayed in Fig.~\ref{fig:AdaNN_vs_NN}. The network structure of NN is the same as that of AdaNN. Before online equalization, both AdaNN and the NN are trained offline using $25\%$ data in set1 (received optical power $-2.7$ dBm). When processing set2, the BER of NN rises to about $1.6 \times 10^{-2}$ abruptly, and remains unchanged since it's non-adaptive. While AdaNN's BER also rises when first encountering set2, the BER soon drops below $10^{-3}$. Note that it only takes about $40$ batches before the BER stabilizes again.
We've also tested another AdaNN model, which is initially trained offline using a different dataset. The ROP of the new dataset is $-4.7$ dBm. Surprisingly, compared with AdaNN (trained@$-2.7$ dBm), AdaNN (trained@$-4.7$ dBm) can achieve very similar BER. This indicates that the adaptive training process of AdaNN is robust even when a different offline-trained model is used. 

\subsubsection{Compare: NN with training sequence}
\begin{figure}[t]
\centering
\includegraphics[width=0.43\textwidth]{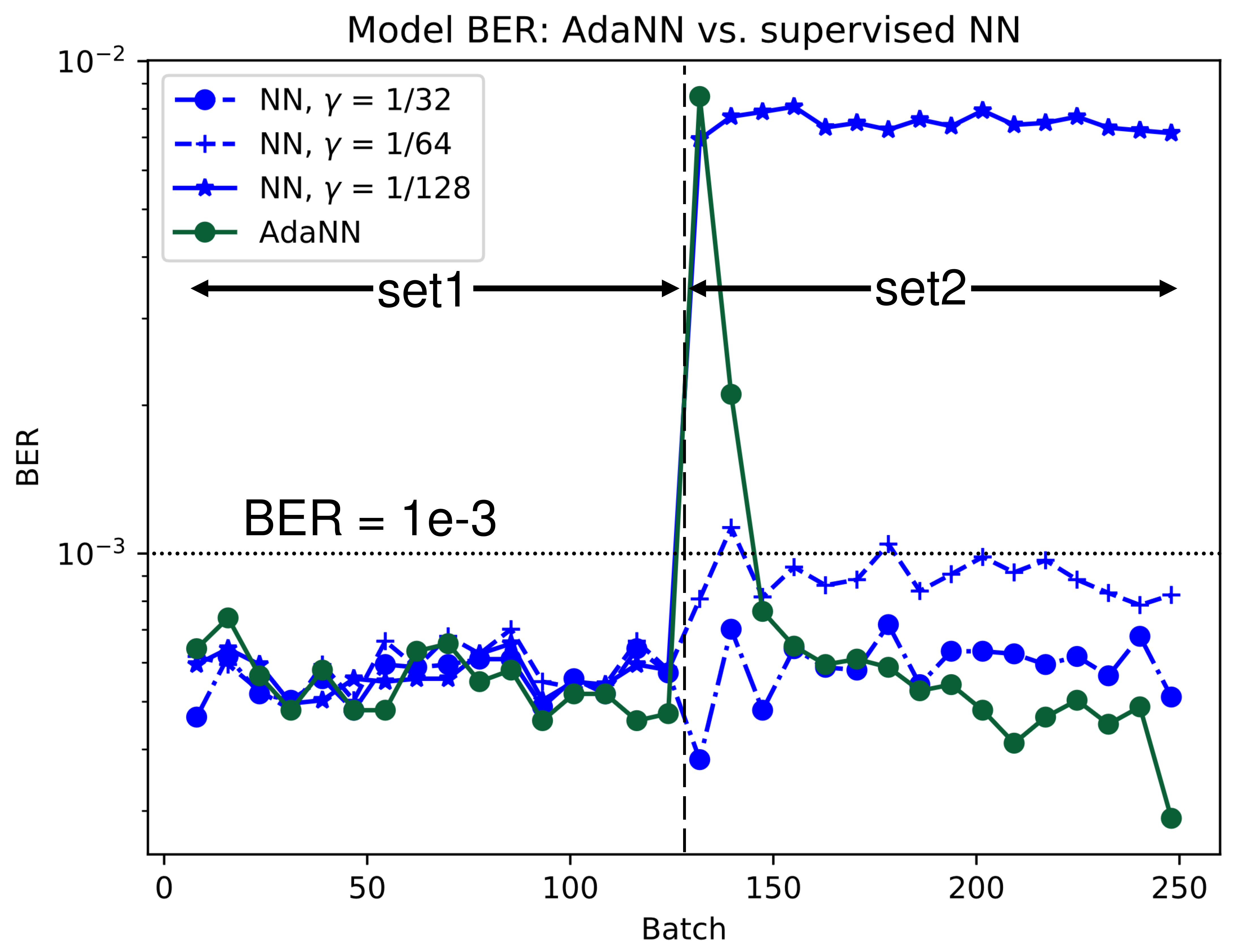}
\caption{The BER performance of normal NN-based equalizer, trained in a supervised manner. $\gamma$ denotes the proportion of training sequence.}
\label{fig:gamma}
\end{figure}
We've already demonstrated that AdaNN can adjust its parameters without the help of labels. It's still necessary to investigate the BER performance of normal NN when short training sequence can be provided. In this part, training sequences are provided at the beginning of set1 and set2 respectively. Concretely, NN's BER on a single batch is measured immediately after NN has been trained on this batch. By minimizing the cross-entropy loss function, the NN is trained using the training sequences for $100$ iterations, ensuring convergence. Denote the ratio of training sequence length to set1 length (or set2 length) as $\gamma$. Fig.~\ref{fig:gamma} shows the BER of NN-based equalizer, fine-tuned with provided training sequence. If $\gamma > 1/32$ (including more than $32768$ symbols), the performance is similar to AdaNN. For smaller $\gamma$, the performance degradation becomes unacceptable. Our results show that AdaNN can achieve better (at least similar) BER performance compared with NN-based equalizer, even when a portion of labels are provided to that NN. There are cases when sending extra training sequence cannot be supported. AdaNN provides an effective alternative for these occasions.

\subsubsection{Compare: Conventional MLSE}
\begin{figure}[b]
\centering
\includegraphics[width=0.43\textwidth]{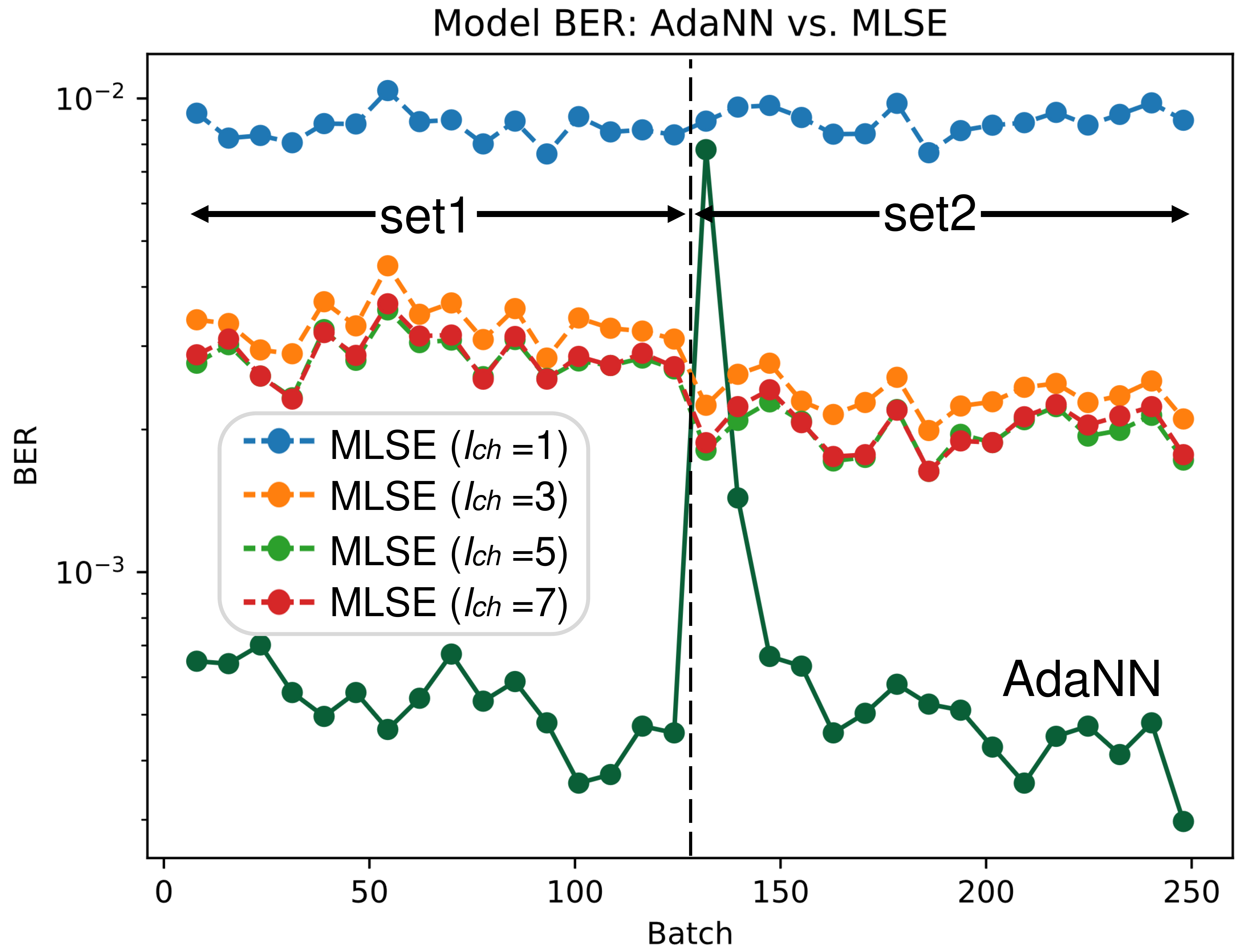}
\caption{The BER performance of AdaNN and conventional MLSE.}
\label{fig:MLSE}
\end{figure}
We have also compared AdaNN with conventional MLSE. The memory length of MLSE takes its value from $l_{ch} \in \{ 1, 3, 5, 7\}$. The channel response coefficients are estimated using least mean square (LMS) algorithm. Note that true labels are provided when updating these coefficients, indicating that MLSE works adaptively in a supervised manner. The update frequency of channel response coefficients is exactly the same as the update frequency of AdaNN parameters. The BER performance of both AdaNN and MLSE are displayed in Fig.~\ref{fig:MLSE}. Obviously, once AdaNN converges it has much lower BER ($3.0\times 10^{-4}$) compared with MLSE ($2.4\times 10^{-3}$). The results show that AdaNN's generalization ability is stronger than adaptive conventional algorithms. 

\subsubsection{Discuss: Influence of Batch Size}
\begin{figure}[!t]
\centering
\includegraphics[width=0.43\textwidth]{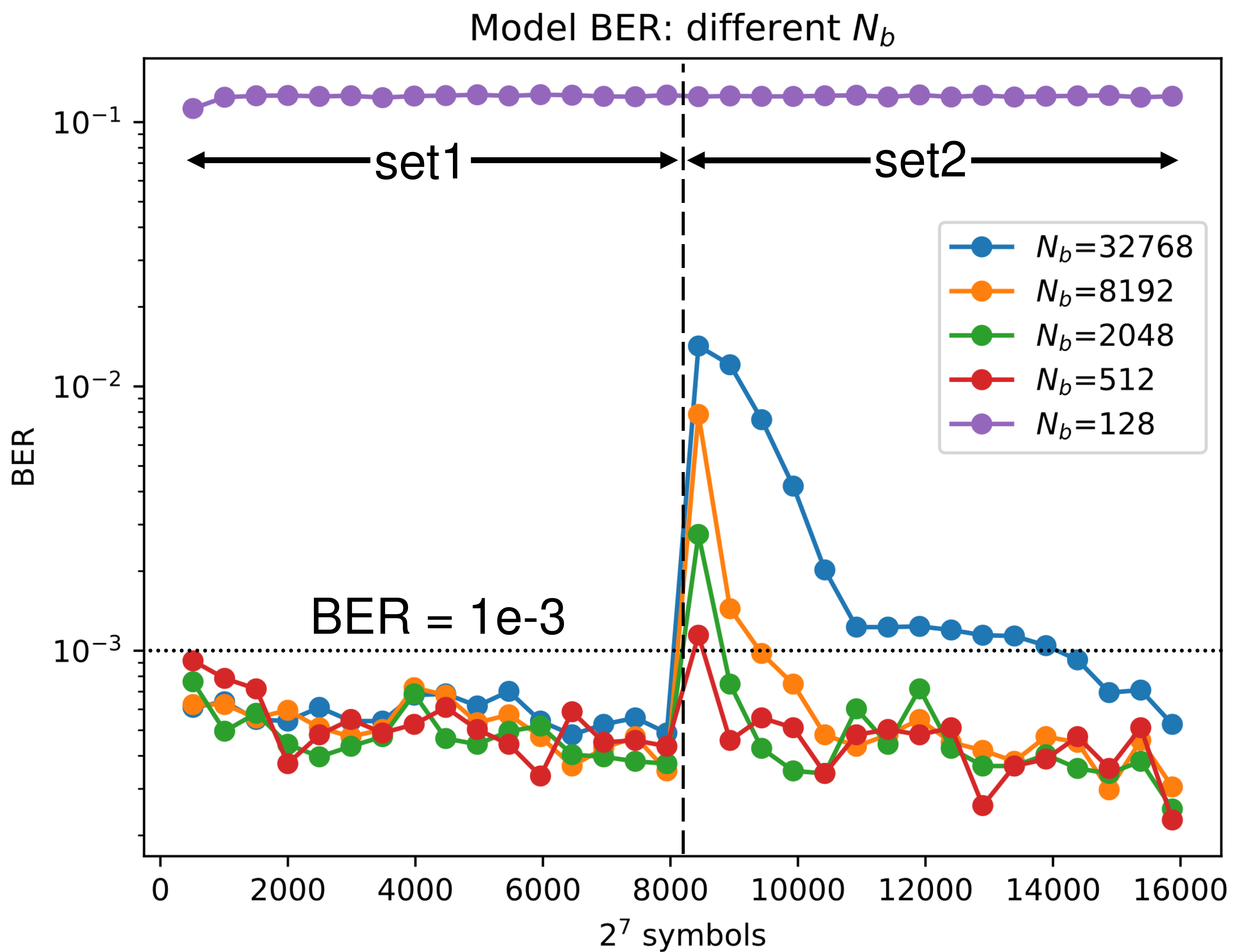}
\caption{The BER performance when choosing different $N_{b}$, following chronological order. Different curves corresponds to different $N_{b}$.}
\label{fig:N_b}
\end{figure}

In previous paragraphs, the batch size $N_{b}=8192$. Choosing different $N_{b}$ will influence the online training process. Concretely, $N_{b}$ describes how much data needs to be collected before AdaNN updates its parameters. Smaller $N_{b}$ seems beneficial since model parameters are updated more frequently. Unfortunately, a very small $N_{b}$ causes new problem, since it leads to very small batch containing few data, which does not reflect the overall probability distribution. Fig.~\ref{fig:N_b} provides the BER performance when $N_{b}$ takes different values, ranging from $128$ to $32768$. As can be seen from the case $N_{b}=128$, when trained on very small batches AdaNN may fail to converge. When building an actual system, $N_{b}$ should be chosen carefully, depending on how frequently the link properties change. 

\begin{figure*}[t]
\centering
\includegraphics[width=0.65\textwidth]{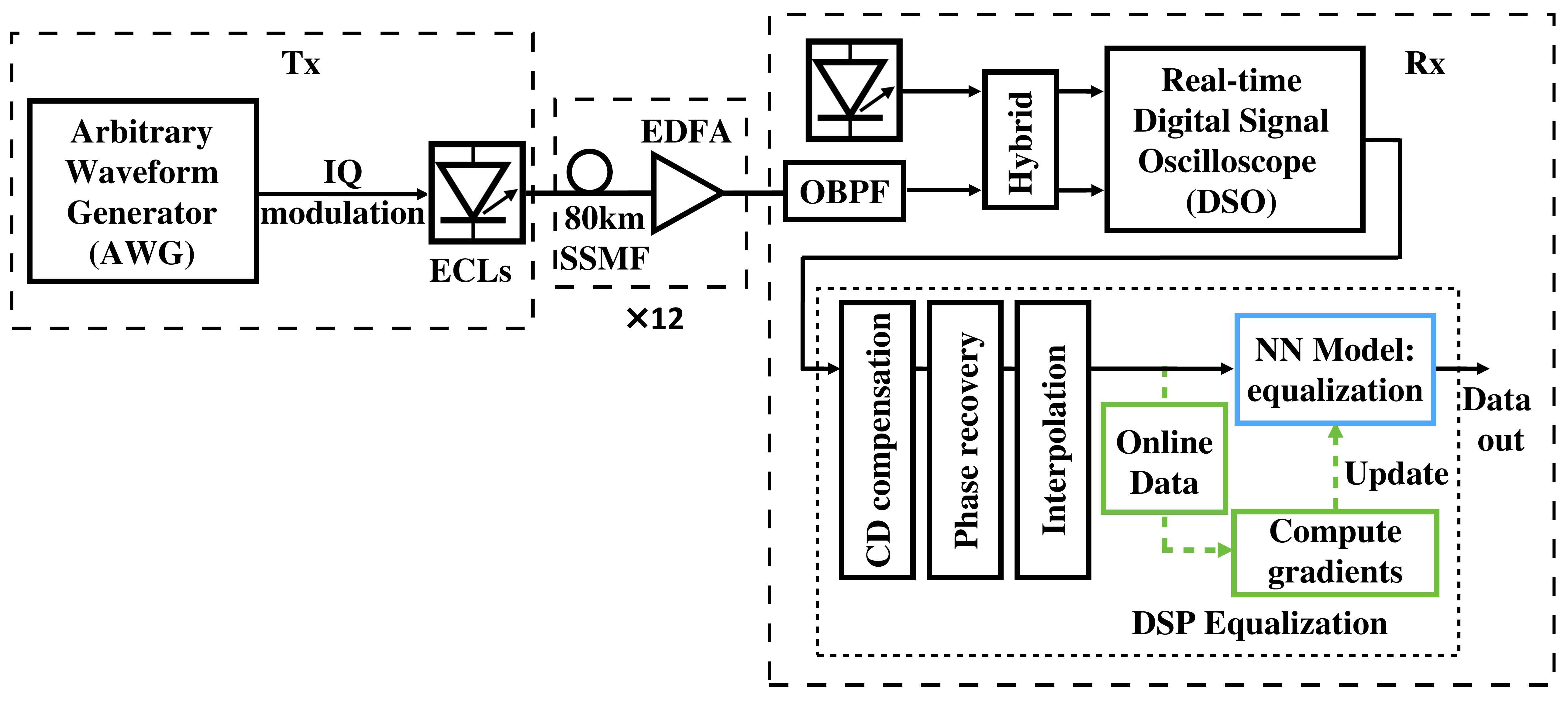}
\caption{Experimental configuration of the 32 Gbaud 16QAM WDM system utilizing ECLs and 960-km SSMF.}
\label{fig:960_system}
\end{figure*}

\subsection{Long-distance optical transmission}
In fact, AdaNN can be used in many communication systems. In this subsection, in order to show its wide applicability, AdaNN is evaluated with a 32 Gbaud 16QAM-modulated Nyquist-WDM system, which is depicted in Fig.~\ref{fig:960_system}. An arbitrary waveform generator operating at 64 GSa/s generates 32 Gbaud 16QAM baseband signals. A root-raised-cosine (RRC) filter with a roll-off factor of $0.1$ is chosen for Nyquist pulse shaping. At the transmitter, we use external cavity laser (ECL) with narrow linewidth of $25$ kHz. The transmission link consists of 12 spans of 80 km SSMF with erbium-doped fiber amplifier (EDFA) only amplification. At the receiver, an optical band pass filter (OBPF) with 45 GHz bandwidth is used as the receiving filter. The coherent receiver consists of an optical local oscillator (LO) with 25 kHz linewidth, optical hybrid, and balanced detectors (BD). A real-time oscilloscope operating at 80 GSa/s stores the received signal. The offline DSP has several stages. First a FIR filter roughly compensates for accumulated dispersion, then carrier frequency recovery is conducted. After synchronization, carrier phase recovery is conducted, and finally AdaNN is used to mitigate nonlinear distortions. 

\begin{figure}[]
\centering
\includegraphics[width=0.43\textwidth]{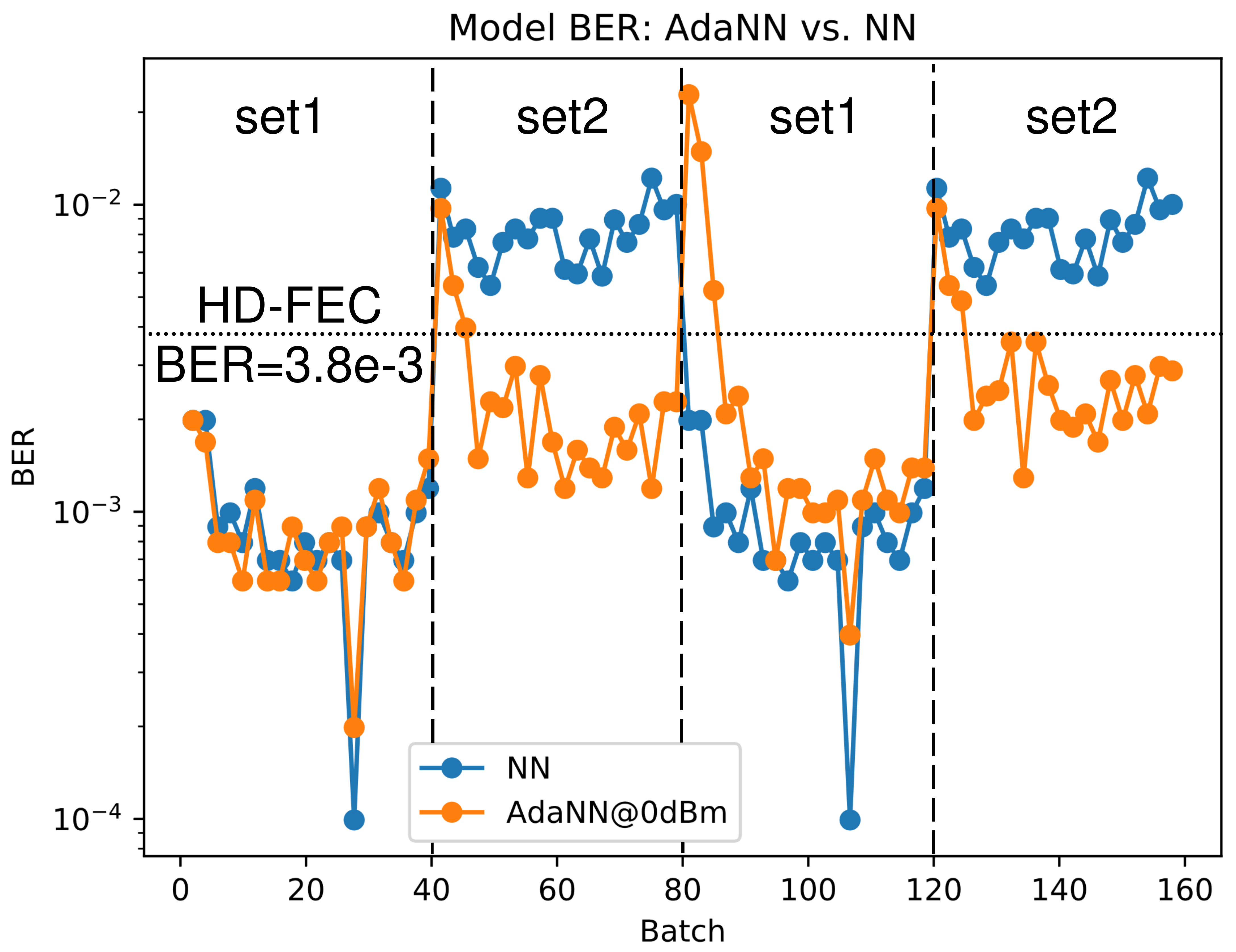}
\caption{The BER performance of a normal NN-based equalizer, compared with AdaNN. After AdaNN stabilizes, its BER on set2 can be lower than $3.8\times 10^{-3}$, which is the $7\%$ HD-FEC limit.}
\label{fig:960}
\end{figure}

Several different sets of 16QAM symbols are collected, each containing $50400$ symbols. Two sets are chosen (denoted as set1 and set2) and concatenated. For set1, the ROP is $0$ dBm, while for set2 the ROP is $-1$ dBm. The polarization of set2 is different from set1. The received signals are 4x resampled ($\Gamma=4$). AdaNN with $4$ hidden layers ($l_{\textrm{NN}}=6$, $R=10$) is first trained offline using $50\%$ data in set1. During online stage, the batch size is $N_b = 1260$. When processing set1 and set2 sequentially, the BER performance on different batches are displayed in Fig.~\ref{fig:960}. Again, AdaNN shows adaptivity and performs better than non-adaptive NN. 

\section{Conclusion}
In this paper, we propose an adaptive online training scheme, which can be used to fine-tune NN-based equalizer without the help of training sequence. 
The proposed adaptive NN-based equalizer is called ``AdaNN''. 
At the online stage, recently received data are collected using a sliding-window. 
With the help of unlabeled data, all the parameters in our NN are fine-tuned in an unsupervised manner, which is similar to decision-directed adaptive equalization.
The performance of AdaNN is evaluated under two scenarios: a 56 Gb/s PAM4-modulated VCSEL optical link, and a 32 Gbaud 16QAM-modulated optical transmission system (960-km SSMF). Heterogeneous datasets are concatenated to test AdaNN's adaptivity. 
Our experimental results indicate that by introducing AdaNN, the BER performance can be improved compared with both non-adaptive NN-based equalizers and conventional MLSE. 
Compared with self-training which serves as a baseline, AdaNN's convergence speed can be 4.5 times faster. 
The online training process has been proved robust when different offline-trained models are used, which shows AdaNN's wide applicability. 
The computational complexity of AdaNN training scheme is also analyzed theoretically. 
We conclude that it is feasible to construct adaptive NN-based equalizer with acceptable computational cost when training sequences aren't provided. 
The generalization ability of all NN-based equalizers can be greatly improved using our proposed method. 


%

\appendices


\section{The complexity of back-propagation}
For AdaNN, all the parameters (including both weights and biases) need to be adjusted online, based on gradients $\nabla_{\boldsymbol{W}_{k}}L_{loss}$ and $\nabla_{\boldsymbol{b}_{k}}L_{loss}$. According to \cite{30}, all these gradients are calculated by implementing back-propagation algorithm, which is given in Algorithm \ref{alg:backward}.
\begin{algorithm}
\caption{Backward propagation}
\label{alg:backward}
\begin{algorithmic}[1]
\Require $l_{\textrm{NN}}$ = the number of layers in the network
\Require $\boldsymbol{W}_{k}, k \in \{ 1, ..., l_{\textrm{NN}}-1\}$ = weight matrices
\Require $\boldsymbol{b}_{k}, k \in \{ 1, ..., l_{\textrm{NN}}-1\}$ = bias vectors
\Require $\boldsymbol{v}$ = the input feature vector

\State $\boldsymbol{g} \gets \nabla_{\boldsymbol{o}} L_{loss}$ $\triangleright$ last layer
\State \textbf{for} ${k} = l_{\textrm{NN}}-1, ..., 1$ \textbf{do}
	\State \quad $\boldsymbol{g} \gets \nabla_{\boldsymbol{a}_{k}} L_{loss} = \boldsymbol{g} \odot \sigma'(\boldsymbol{a}_{k})$
	\State \quad $\nabla_{\boldsymbol{b}_{k}} L_{loss} = \boldsymbol{g}$
	\State \quad $\nabla_{\boldsymbol{W}_{k}} L_{loss} = \boldsymbol{g} \boldsymbol{h}_{k-1}^{\top}$
	\State \quad $\boldsymbol{g} \gets \nabla_{\boldsymbol{h}_{k-1}} L_{loss} = \boldsymbol{W}_{k}^{\top}\boldsymbol{g}$ $\triangleright$ continue to layer $k-1$
\State \textbf{end for}
	
\end{algorithmic}
\end{algorithm}

Consider the back-propagation from layer $k$ to layer $k-1$. The number of neurons contained in each layer  can be denoted as $R_{k}$ and $R_{k-1}$. It's straightforward to see that, during the back-propagation from layer $k$ to layer $k-1$, $\boldsymbol{g} \gets \boldsymbol{g} \odot \sigma'(\boldsymbol{a}_{k})$ requires $R_{k}$ multiplications, $\nabla_{\boldsymbol{W}_{k}} L_{loss} = \boldsymbol{g} \boldsymbol{h}_{k-1}^{\top}$ requires $R_{k-1} \cdot R_{k}$ multiplications, and $\boldsymbol{g} \gets \boldsymbol{W}_{k}^{\top}\boldsymbol{g}$ requires $R_{k-1} \cdot R_{k}$ multiplications. By summing over all layers, we can conclude that for a single back-propagation, the number of required floating-point multiplications $k_{\textrm{back}}$ can be calculated as
\begin{equation}
\begin{split}
k_{\textrm{back}} =&~[2\Gamma(2L+1)\cdot R + R] \\
&+ (l_{\textrm{NN}}-3)\cdot (2R^{2} + R) + (2R\cdot M + M) \\
=&~ 2k_{\textrm{NN}} + (l_{\textrm{NN}}-2)\cdot R + M.
\end{split}
\end{equation}

\section{Weight changes}
It would be useful to know what changes does online training stage have on NN parameters. We've compared the weight matrices in two models: the initial model trained offline, and the final model which has been online trained with set1 and set2. Our NN model has $l_{\textrm{NN}}=6$ layers: $1$ input layer, $4$ hidden layers, and $1$ output layer. There are $5$ weight matrices between these $6$ layers, denoted as $\boldsymbol{W}_{1}, ..., \boldsymbol{W}_{5}$.

Define function $S(\boldsymbol{W})$ as summing up the absolute value of all elements in matrix $\boldsymbol{W}$:
\begin{equation}
S(\boldsymbol{W}) = \sum_{i,j} |W_{ij}|.
\end{equation}
We now calculate the following ratio for all layers ($k=1,2,...,5$):
\begin{equation}
r_{k} = \frac{S(\Delta \boldsymbol{W}_{k})}{S(\boldsymbol{W}^{\textrm{init}}_{k})} = \frac{S(\boldsymbol{W}^{\textrm{final}}_{k} - \boldsymbol{W}^{\textrm{init}}_{k})}{S(\boldsymbol{W}^{\textrm{init}}_{k})}.
\end{equation}
The ratio $r_{k}$ reflect the change of weight matrix $\boldsymbol{W}^{\textrm{init}}_{k}$. The results are given in Table.~\ref{tab:r_k}.

\begin{table}[!h]
	\caption{Ratio $r_{k}$ for weight matrices between different layers.}
	\label{tab:r_k}
	
	\centering
	\begin{tabular}{cccc}
	    \hline
		\hline
		$k$ & $S(\boldsymbol{W}^{\textrm{init}}_{k})$ & $S(\Delta \boldsymbol{W}_{k})$ & $r_{k}$ \\
		\hline
		\hline
		$1$ & $71.122$ & $5.788$ & $0.081$ \\
		\hline
		$2$ & $30.641$ & $0.929$ & $0.030$ \\
		\hline
		$3$ & $31.368$ & $0.250$ & $0.008$ \\
		\hline
		$4$ & $34.080$ & $0.189$ & $0.006$ \\
		\hline
		$5$ & $21.338$ & $0.172$ & $0.008$ \\
		\hline
		
	\end{tabular}
\end{table}

It can be concluded that only minor changes have occurred during online training stage. On the other hand, it can be observed that $r_{k}$ becomes larger for weight matrices near the input layer.

\ifCLASSOPTIONcaptionsoff
  \newpage
\fi

\end{document}